\documentclass[aps,epsf,preprint]{revtex4}
\usepackage{amssymb}
\usepackage{amsmath}
\usepackage{color}

\def\[{\left[}
\def\]{\right]}
\def\be{\begin{eqnarray}}
\def\ee{\end{eqnarray}}
\def\ol m_0{\begin{pmatrix}}
\def\bm{\begin{pmatrix}}
\def\em{\end{pmatrix}}
\def\ba{\begin{array}}
\def\ea{\end{array}}
\def\bi{\begin{itemize}}
\def\ei{\end{itemize}}

\def\({\left(}
\def\){\right)}

\def\eq#1{Eq.(\ref{#1})}
\def\a{\alpha}
\def\s{\sigma}
\def\e{\epsilon}

\def\f{\phi}

\def\m{\mu}
\def\w{\omega}

\def\n{\nu}
\def\p{\partial}
\def\d{\delta}
\def\labels#1{\label{#1}}
\def\bn{\begin{enumerate}}

\def\en{\end{enumerate}}
\def\b{\beta}
\def\g{\gamma}
\def\ba{\begin{array}}
\def\ea{\end{array}}
\def\bc{\begin{center}}
\def\ec{\end{center}}

\def\ni{\noindent}
\def\.{\!\cdot\!}
\def\igw#1{\includegraphics[width=#1cm]}
\def\igwg#1#2{\igw{#1}{#2.png}} 
\def\+{\!+\!}
\def\-{\!-\!}
\def\vs{\vskip.5cm}

\def\r{\rho}
\def\h{{1\over 2}}
\def\Rw{\Rightarrow}
\def\={\stackrel{.}{=}}

\def\ol{\bar}
\def\bn{\bar n}
\def\1{\bar 1}
\def\8{\bar 8}
\def\9{\bar 9}
\def\7{\bar 7}

\def\nr{\nonumber}
\def\t{\tau}

\def\bba{\bar\a}
\def\bbm{\bar m}
\def\bbr{\bar\r}
\def\bbp{\bar p}
\def\bbu{\bar u}

\def\bbe{\bar\e}

\usepackage{amssymb}
\usepackage{graphicx}

\begin{document}

\title{Dynamical Property of Black Hole Matter}
\author{C.S. Lam}
\affiliation{Department of Physics, McGill University\\
 Montreal, Q.C., Canada H3A 2T8\\
Department of Physics and Astronomy, University of British Columbia,  Vancouver, BC, Canada V6T 1Z1 \\
Email: Lam@physics.mcgill.ca}

\begin{abstract}
Matter loses its original characteristics after entering a black hole, thus
becoming a new kind of (black hole) matter. The property of this new matter cannot
be measured experimentally, but some of it can be deduced theoretically from the
Einstein equations and the conservation laws which it must still satisfy. In a previous paper, this matter is modelled by an ideal fluid, with an equation of state
$p(r)=-\xi\r(r)$ between the pressure $p(r)$ and the density $\rho(r)$. In order for
this matter to fill the inside of a black hole so that its property can be teased out
from the Einstein and conservation equations, it must possess a negative pressure
($\xi>0$) to counter the gravitation attraction which draws all matter
to the center. In that case a solution of the Einstein and conservation equations exists if and only if the constant $\xi$ is confined within a narrow range, between 
0.1429 and 0.1716. In the present paper, we try to find out its dynamical response
 by injecting additional matter into the black hole over a period of time. The resulting solutions of the six time-dependent Einstein equations and conservation
laws are presented in perturbation theory, valid if the total amount of injection
is small. Even in perturbation, the solutions can be obtained only with a special trick. The result shows that the equation of state $p(r,t)=-\xi\r(r,t)$ remains unchanged
with the same $\xi$  when the injection rate is constant. When the rate changes
with time, $\xi$ requires a correction,
$\xi\to\xi+\xi_1(r,t)$, where $\xi_1(r,t)$
appears to be correlated with the acceleration of the injected matter in a way to be shown in the text.
\end{abstract}
\maketitle

\section{Introduction}
According to the no-hair theorem \cite{MTW73,sI67,sI68}, a stationary black hole is completely
characterized by its mass, its electric charge, and its spin. For  neutral and non-rotating
black holes, the only thing that tells them apart is their mass. In the Standard
Model, other than mass and electric charge, a particle is  characterized also  by its baryonic number, isotopic spin, hypercharge, and statistics, etc. With the no-hair theorem, all these  are lost once matter enters into the black hole, so what is inside
a black hole must be a new kind of matter, to be referred  hereafter as the `black hole matter'. Very little is known about it because there is no way to
get inside the black hole to measure. Nevertheless, it must obey the Einstein
equations and the conservation laws, from which some information of the new matter
can be teased out. There are some studies along this line \cite{MM15, BMN19, MM23} but to my knowledge
not many. The most popular and well known model is the Schwarzschild solution,
which assumes all black hole matter sink to the singularity at the center of the black hole, leaving a vacuum prevailing everywhere else inside the black hole. This solution unfortunately tells us nothing about the nature of the new matter as
it is nowhere to be found except at the singularity.

In a recent paper \cite{Lam}, we studied the black hole matter inside a neutral, spherical, and non-rotational black hole modelled by an ideal fluid, whose density $\r(r)$
and pressure $p(r)$ are related by the equation of state $p(r)=-\xi\r(r)$. In order for
matter to fill the black hole so that its property can be probed,
a negative pressure ($\xi>0$) is required to counteract the gravitational attraction  which tends to draw everything to the singularity at the center. In order to satisfy the
Einstein equations and the conservation laws, we found that the constant parameter $\xi$
must be confined to a narrow range, between $\xi=1/7\simeq 0.1429$ and  $\xi=3-2\sqrt{2}\simeq 0.1716$. A singularity of $\r(r)$ still appears at the origin
as demanded by the singularity theorem \cite{aK56,aR57,rP65,SG14, kL22}, but $\r(r)$ is also non-zero elsewhere
inside the black hole, except at the event horizon where it must vanish
to match the vacuum condition outside the black hole.

In this paper, we try to find out more property of the new matter by probing its
dynamical response. About the only thing one can do to  probe its reaction
is to inject more matter into the black hole at different rates. 
 In this way, starting from a static black hole of mass $M_0$, 
additional mass is injected to increase it from $M_0$ to $M(t)=M_0(1+\s(t))$ at time $t$.
 By calculating $\r(r,t)$ and $p(r,t)$ from the Einstein equations and the conservation laws
for an arbitrary $\s(t)$, one can in principle deduce how black hole responds to such a dynamical probing.

Unfortunately, the resulting Einstein and conservation equations (Appendix A) are so complicated
and so nonlinear that there is no hope to find an exact solution. For that reason we
turn to perturbative solutions by assuming $\s(t)\ll 1$ at all times. With that caveat and a trick to be explained, the equations
can then be solved to obtain  $\r(r,t)$ and $p(r,t)$ and other quantities. 

We find that the static equation of state $p(r,t)=-\xi\r(r,t)$ remains unchanged with
the original $\xi$ if the second time derivative $\ddot\s(t)=0$.  If $\ddot\s(t)\not=0$,
then the constant $\xi$ receives a correction, $\xi\to\xi+\xi_1(r,t)$, with $\xi_1(r,t)=\hat\xi_1(r)\ddot\s(t)$. 

That the equation of state can be altered by external conditions is also seen in ordinary matter. A non-relativistic gas has an equation of state with $\xi=0$, but
if heat is applied to that system to make the movement relativistic, then $\xi=-{1\over 3}$. Similarly, in the present case changes of equation of state occurs if the injected matter has an acceleration $\ddot\s(t)\not=0$. 

Sec.~II reviews the black hole matter model proposed in \cite{Lam}. It is  the
zeroth order solution of the time-dependent Einstein equations and  conservation laws
to be discussed in Sec.~III. 
The first-order equation are shown in \eq{CO1x} to \eq{EQ33x}, but it is not easy
solve them directly. However, their time-independent version can be solved
using a trick discussed in Sec.~IV, which then serves as an effective stepping stone
to yield the solutions of the time-dependent equations to be presented in Sec.~V.
These solutions are summarized in Sec.~VIA, their validity and interpretations
discussed in Sec.~VIB, and the solutions are illustrated numerically in Sec.~VIC.

\section{The Static Model}
The matter distribution inside a neutral, non-rotating, and spherical black hole is reviewed here.  In this model \cite{Lam}, black hole matter is described by an ideal fluid with the energy-momentum tensor
\be T_{\m\n}=(p+\r)U_\m U_\n+p g_{\m\n}, \labels{semt}\ee
where as usual the four-velocity is  normalized to $U^\m U_\m=-1$. The energy density $\r$ is positive,
and the pressure $p$ obeying the equation of state $p=-\xi\r$ is negative.
The constant $\xi$ has to be confined to a narrow range
$0.1429<\xi<0.1716$  in order to satisfy the Einstein equation and
the covariant conservation laws.

The spacetime metric
\be ds^2&=&-e^{2\a(r)}dt^2+\(1-2Gm(r)\over r\)^{-1} dr^2+r^2 d\Omega^2\labels{ds2}\ee
is Schwarzschild  outside the black hole, where $e^{2\a(r)}=\(1-{2GM\over r}\)$ and $m(r)=M$, with $G$ being the gravitational constant and
$M$  the black hole mass with an event horizon located at $r=R=2GM$. Inside the black hole, the metric is no longer
Schwarzschild because it is now filled with matter given by \eq{semt}. 
Einstein equations of this system lead to the Tolman-Oppenheimer-Volkoff equation
for the pressure gradient  \cite{sC04, OV39}
\be {dp\over dr}&=&-{(\r+p)[Gm(r)+4\pi Gr^3p]\over r[r-2Gm(r)]},\labels{dpdr}\ee
and a relation between $m(r)$ and $\r(r)$,
\be m(r)&=&4\pi\int_0^r\r(r'){r'}^2dr',\labels{massr}\ee
so $m(r)$ can be interpreted as the mass of a ball of  black-fluid with a radius $r\le R$.
The  metric function $\a$  inside the black hole satisfies the equation
\be {d\a\over dr}&=&-{1\over (\r+p)}{dp\over dr}.\labels{dadr}\ee

Continuity across the black hole surface  is imposed as the boundary condition
to match the inside and the outside of the black hole:
\be p(R)=\r(R)=0,\quad m(R)=M,\quad e^{2\a(R)}=0.\labels{bondc}\ee

For $R-r$ small and positive, \eq{dpdr}
can be approximated by
\be {dp\over dr}=-{(\r+p)GM\over R[r-2GM]}={\xi-1\over 2\xi}{p\over R-r},\quad (R-r\ll R),\labels{dpdrR}\ee
whose solution for small and positive $R-r$ is
\be p(r)\simeq -\tilde c(R-r)^{(1-\xi)/2\xi}\equiv -\tilde c(R-r)^\g \labels{pR}\ee
for some $\tilde c>0$. Thus $0<\xi<1$ is required in order to have $p(R)=0$, so the black-hole matter must carry
a negative pressure. 

Using \eq{dadr} and $p=-\xi\r$, one gets 
\be e^{2\a(r)}=c_1\r(r)^{1/\g}\labels{e2a}\ee
for some $c_1>0$. By rescaling  time $t$ in \eq{ds2}, one can make $c_1=1$.
When $r$ approaches
$R$, \eq{pR} shows that $e^{2\a(r)}\to c_2(R-r)$ for some $c_2>0$,  matching the value 0 for this metric component
outside the black hole.

These equations can be written in a dimensionless form by letting
let $x=r/R$, $\bbm(x)=m(r)/M$, $\bbr(x)=\r(r)(R^3/M)$,  $\bbp(x)=p(r)(R^3/M)=-\xi\bbr(x)$, and $\bba(x)=\a(r)$. Then \eq{dpdr} in the dimensionless form becomes
\be
{d\bbr(x)\over dx}&=&-{\xi-1\over 2\xi}{\bbr(x)\over x}{\bbm(x)-4\pi\xi x^3\bbr(x)\over x-\bbm(x)},\labels{dless1}\\
{d\bbm(x)\over dx}&=&4\pi x^2 \bbr(x), \labels{dless2}\ee
and the boundary conditions are $\bbr(1)=0$ and $\bbm(1)=1$. \eq{dadr} can be written as
\be {d\bba(x)\over dx}={\xi\over 1-\xi}\({1\over\bbr(x)} {d\bbr(x)\over dx}\),\labels{dless3}\ee
whose solution from \eq{e2a} is
\be e^{2\bba(x)}=\bbr(x)^{1\over\g}.\labels{dless4}\ee

The interior solution equivalent to \eq{pR} near $x=1$ is then
\be
\nr\bbr(x)&\simeq& c_3(1-x)^{(1-\xi)/2\xi}\equiv c_3(1-x)^\g,\\
\nr\bbm(x)&\simeq&1-{8\pi \xi c_3\over \xi+1}(1-x)^{(\xi+1)/2\xi}=1-{8\pi \xi c_3\over \xi+1}(1-x)^{\g+1},\\
e^{2\ol\a(x)}&\simeq& c_3^{1/\g}(1-x)\qquad (1-x\ll 1)\labels{solrm}\ee
for some $c_3>0$. 

Near $x=0$, $\bbm(x)=\m_0x^\b$ for some $\m_0>0$, and it is necessary to impose $0<\b<1$ in order for \eq{dless1} and \eq{dless2} to have a solution. It follows from \eq{dless1} that $\bbr(x)=(\m_0\b/4\pi) x^{\b-3}$, and  $\b$ is related to $\xi$ by the relation
\be \b=(\-1 \+ 7 \xi)/\xi (1 \+ \xi).\labels{beta}\ee
The requirement for $\b$ to be between 0 and 1 forces $\xi$ to be in a narrow range between
$\xi=1/7\simeq 0.1429$ (when $\b=0$) and  $\xi=3-2\sqrt{2}\simeq 0.1716$ (when $\b=1$).

The corresponding values for $\g=(1-\xi)/2\xi$, the power of $\ol\r(x)$ near $x=1$, are $\g=3$ when $\xi=1/7$, and
$\g=2.414$ when $\xi=3-2\sqrt{2}$.

A numerical solution for $\xi=0.16$ is shown in Fig.~1.
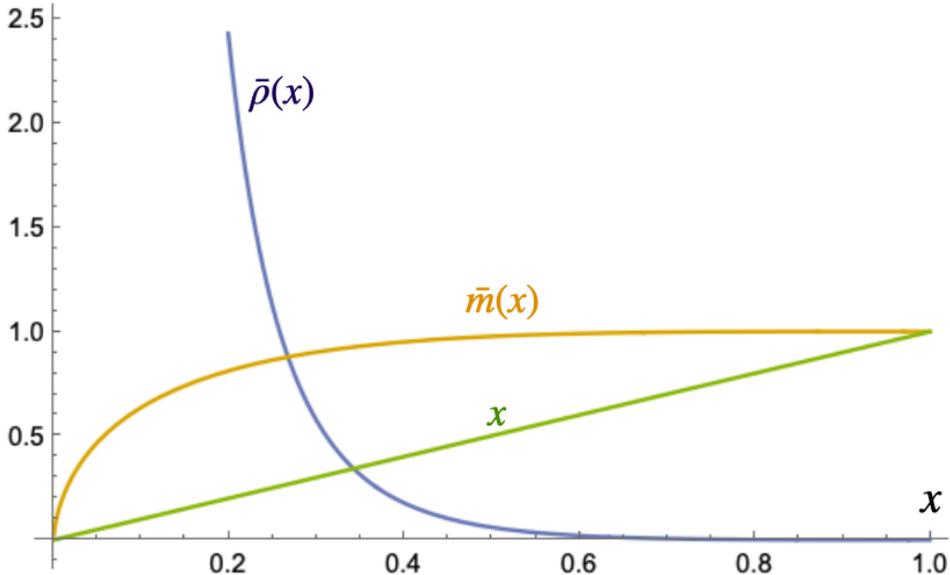
\bc\igwg{14}{Fig1}\\ Fig.~1.\quad Scaled mass and density distribution in the black-hole interior for $\xi=0.16$\ec

This diagram illustrates the genereal behavior of functions true for any $\xi$ within the  allowed range.  $\bbr(x)$ decreases steadily from $\sim x^{\b-3}\ (\b<1)$ at small $x$ towards $\sim(1-x)^\g\ (\g>0)$ near $x=1$. In the mean time, $\bbm(x)$ increases steadily from $\sim x^\b$ at small $x$ towards 1, keeping $\bbm(x)-x\ge 0$.
The steady increase of $\bbm(x)$ is a consequence of \eq{dless2} and $\bbr(x)\ge0$,
and the positivity of $\bbm(x)-x$ except at the boundaries is needed
to avoid a singularity appearing in \eq{dless1}. 

\section{Time-dependent injection} 
We found in the last section that black hole matter can be represented by an ideal fluid with an equation of state
$p(r)=-\xi\r(r)$, as long as $0.1429<\xi<0.1716$. In the rest of this article, we  shall investigate the dynamical property
of this matter, deduced from its response to an  injection of additional matter into the black hole.
As the response cannot be measured experimentally, we must resort to the solution of time-dependent Einstein equations to find out what it is.

Intuition might tell us that the internal adjustment to such an input would start at the surface of the black hole, then  propagating inward 
 at the speed of sound. However, such an intuition based on physics of compressible fluid in a spacetime with a signature
$(-,+,+,+)$ may not be applicable inside a blackhole whose signature is $(-,-,+,+)$. After all, with this change of signature,
a hyperbolic wave equation in $r$ and $t$ becomes an elliptic equation, suggesting that the internal adjustment might  actually be instantaneous.

The only way to find out the truth is to solve the time-dependent Einstein equations, but unfortunately that proves to be too difficult a task.
To make some headway, we shall consider the idealized situation where matter is accreted spherically,
with a total amount so small that perturbation theory can be applied to solve the Einstein equations.

Time-dependent density $\r$ and pressure
$p$ require  time-dependent metric functions $\a$ and $m$ in \eq{ds2t}. The continuous injection of matter brings about
a fluid flow with a non-zero momentum density  $T_{tr}$, which in turn  requires a non-diagonal metric element $g_{tr}$. The appropriate metric to be used 
inside the black hole is therefore
\be ds^2&=&-e^{2\a(r,t)}dt^2+\(1-{2Gm(r,t)\over r}\)^{-1}dr^2+2\e(r,t)drdt+r^2 d\Omega^2\equiv g_{\m\n}dx^\m dx^\n.\labels{ds2t}\ee
As in the stationary case,   $m(r,t)$ can be interpreted as the mass
of a ball of fluid with radius $r$ at time $t$. The event horizon $R(t)$ at time $t$ is related to the black hole mass $M(t)$
by $R(t)=2GM(t)$, where $M(t)=m(R(t),t)$. 

The time-dependent energy-momentum tensor of an ideal fluid is now
\be T_{\m\n}=\big(\r(r,t)+p(r,t)\big)U_\m U_\n+p(r,t)g_{\m\n}, \labels{SEMT}\ee
where the four-velocity $U^\mu=(u^0(r,t), u(r,t), 0,0)$ [for the components $\m=(t,r,\theta,\f)$] is as usual normalized to $U^\m U_\m=g_{\m\n}U^\m U^\n=-1$. 
This implies 
\be
u^0(r,t)=e^{-2 \alpha (r,t)} \Bigg(u(r,t)\epsilon (r,t)+\sqrt{ e^{2 \alpha (r,t)} \left(\frac{r
u(r,t)^2}{r-2 G m(r,t)}+1\right)+u(r,t)^2 \epsilon (r,t)^2}\Bigg).\labels{u0}
\ee

The computed expressions for the conservation law $\nabla_\m T^\m_{\ \n}=0$ [to be referred to as $CO_\n$] and the Einstein equation $G_{\m\n}-8\pi GT_{\m\n}=0$  [to be referred to as $EQ_{\m\n}$]  are given in Appendix A.
There are only four independent Einstein equations, $EQ_{tt}, EQ_{rr}, EQ_{tr}$, and $EQ_{\theta\theta}$,
because $EQ_{\f\f}$ is the same as $EQ_{\theta\theta}$, and all  other components are zero. There are two  non-zero conservation laws, $CO_t$
and $CO_r$.
In principle, these six equations would be used to solve 
the unknown variables in the metric and the momentum tensor. However, a glance at
Appendix A  makes it obvious that these equations
are so complicated and so non-linear that there is no hope to solve them exactly, 
 so we resort to perturbation theory. Starting from
 a static black hole of mass $M_0$ and radius $R_0$, as described in Sec.~II, 
 additional matter is injected  over a period of time so that the black hole mass at any
 given time becomes $M(t)=M_0(1+\s(t/R_0))$, where $\s(t/R_0)$ is a non-negative and monotonically
 increasing function with $\s(t/R_0)\ll1$ at all time.
 Accordingly, we make the expansions
\be
\a(r,t)&=&\a_0(r)+\a_1(r,t),\\
p(r,t)&=&p_0(r)+p_1(r,t),\\
\r(r,t)&=&\r_0(r)+\r_1(r,t),\\
m(r,t)&=&m_0(r)+ m_1(r,t),\labels{expansion}\ee
and keep only the 0th and 1st order terms of $\e(r,t), u(r,t),\a_1(r,t),  p_1(r,t),\r_1(r,t), m_1(r,t)$ in the equations. 
To this order, it follows from \eq{u0} that $u^0=e^{-\a_0(r)}\(1-\a_1(r,t)\)$.

\subsection{Zeroth order} 
The 0th order equations describe a static black hole that is already considered in Sec.~II. In that case $EQ_{tr}$ is identically zero and the metric is diagonal.
It turns out that $CO_t$ is also zero, and that $EQ_{\theta\theta}=0$ if the following three equations are satisfied. The remaining  equations can be extracted from the expressions in Appendix A to be 
\be
CO_r&&0={d p_0(r)\over dr}+\big(p_0(r)+\r_0(r)\big) {d\a_0(r)\over dr}\labels{A20}\\
EQ_{tt} &&0={dm_0(r)\over dr} -4 \pi  r^2 \rho_0 (r)\labels{A30}\\ 
EQ_{rr} &&0= {d\alpha_0(r)\over dr}r\big(r-2 G m_0(r)\big)-G \big(m_0(r)-4 \pi    r^3 p_0 (r)\big).\labels{A50} \ee
These are equivalent to  \eq{dadr}, \eq{massr}, \eq{dpdr} used to calculate the static black hole if we insert the equation
of state $p_0(r)=-\xi\r_0(r)$,
except that a subscript 0 is now added to the functions to emphasize its static nature.

\subsection{First order} 
To save
writing, arguments of  functions will not be shown below. All first-order functions ($\e, u$ and those with subscript 1) depend on $r$ and $t$, and all
the zeroth order functions (those with subscript 0) depend only on $r$.
Time derivative $\p/\p t$ is indicated by a dot on top, and spatial derivative $\p/\p r$ is indicated by a prime. 

The first-order relations of the six equations can be extracted from the expression in Appendix A. After using
\eq{A20} to \eq{A50} to simplify, the result is 
\be
\ni CO_t&&\labels{CO1x}\\ 
\nr 0&=&\xi  r \bigg(e^{ \a_0 } (r-2 G m_0 )
  \dot\r_1 -(\xi -1)
   \r_0  \Big(e^{2\a_0} (r-2 G
   m_0 ) u' +G
  e^{ \a_0 }
   \dot m_1 \Big)\bigg)\\
\nr &+&(\xi -1) e^{2 \a_0 }
  \r_0  u  \Big(G (4 \xi
   -1) m_0 +2 \xi  r \left(2 \pi  G
   \xi  r^2 \r_0 -1\right)\Big)\\
\ni CO_r&&\labels{CO2x}\\   
\nr 0&=&r e^{2 \a_0 } (r-2 Gm_0 ) \Big(p_1' +(1-\xi)\r_0\a_1'\Big) +G
   e^{2 \a_0 }
    \left(m_0 -4 \pi 
   \xi  r^3 \r_0 \right)( p_1+\r_1)\\
\nr &+& (1-\xi)r \r_0  (r-2 G
   m_0) \dot \epsilon  +(1-\xi ) \left(e^{ \a_0 }\right) r^2 \r_0 \dot u \\
\ni EQ_{tt}&&\labels{EQ11x}\\
\nr 0&=&4 \pi  r^2 \r_1-m_1'\\
\ni EQ_{tr}&&\labels{EQ12x}\\
\nr 0&=&\dot m_1-4 \pi  (\xi -1) r^2e^{ \a_0}\r_0 u\\
\ni EQ_{rr} &&\labels{EQ22x}\\
\nr 0&=&-4 \pi  G r^2 (r-2 Gm_0)p_1
  +(r-2 G m_0)^2( \a_1'+e^{-2 \a_0}\dot\e)+G
  \left(8 \pi  G \xi  r^2 \r_0-1\right)m_1\\
\ni EQ_{\theta\theta}&&\labels{EQ33x}\\
   \nr 0&=&G    \bigg(32 \pi ^2 G^2 \xi 
   r^5 \r_0   ^2-G m_0   
   \Big(8 \pi  G (4 \xi -1) r^2
  \r_0   +1\Big)+4 \pi  G (3
   \xi -2) r^3 \r_0   +r\bigg)m_1\\
\nr &-&G (2 G m_0   -r)
    \left(G
   m_0   +4 \pi  G \xi  r^3
   \r_0   -r\right) m_1' +G r^2 e^{-2
  \a_0   } (2 G
  m_0   -r) \ddot m_1\\
 \nr  &-&8
   \pi  G r^2 (r-2 G m_0  )^2
  p_1  -e^{-2 \a_0 } (r-2 G m_0   )^2 
     \left(G m_0  +4 \pi
    G r^3 \r_0   -r\right)\dot \epsilon\\
 \nr &-&(r-2
   G m_0   )^2   \Big(-G m_0   +4
   \pi  G (2 \xi +1) r^3 \r_0  -r\Big)\a_1'+r e^{-2 \a_0  } (r-2 G m_0   )^3 \dot \epsilon'\\
\nr    &+&r (r-2 G m_0   )^3
  \a_1''  
  \ee
  
 As in Sec.~II, it is useful to put these equations into a dimensionless form. With the following definitions,
\be
\nr r&=&xR_0,\quad t=\tau R_0,\quad \e(r,t)=\ol\e(x,\tau),\quad \a_0(r)=\ol\a_0(x),
\quad \a_1(r,t)=\ol\a_1(x,\tau)\\
\nr u^0(r,t)&=&\ol u^0(x,\tau)=e^{-\bba_0(x)}\(1-\bba_1(x,\tau)\),\quad u(r,t)=\ol u(x,\tau),\quad p_1(r,t)=\ol p_1(x,\tau)/(2GR_0^2),\\
\nr \r_0(r)&=&\ol\r_0(x)(M_0/R_0^3)=\ol\r_0(x)/(2GR_0^2), \ \ \r_1(r,t)=\ol\r_1(x,\tau)/(2GR_0^2),\\
m_0(r)&=&\ol m_0(x)M_0=\ol m_0(x)(R_0/2G) , \ \ m_1(r,t)=\ol m_1(x,\tau) (R_0/2G),\labels{scaledvar}\ee
and $\bbr_0(x)$  replaced by $\bbm_0'(x)/4\pi x^2$  per \eq{dless2},
 \eq{CO1x}-\eq{EQ33x}  become 
\be    
\ni CO_t &&\labels{CO1xs}\\
\nr 0&=&-(\xi -1) \xi  x \bbm_0'
   \dot m_1-2 (\xi -1)
   \xi  x (x-\bbm_0)e^{
   \bba_0} \bbm_0'
  \bbu'\\
\nr &+&(\xi -1)
  e^{ \bba_0}
   \bbm_0' 
   \Big(\xi  x \left(\xi 
   \bbm_0'-4\right)+(4 \xi -1)
  \bbm_0\Big)\bbu+8 \pi  \xi  x^3
   (x-\bbm_0) \dot\bbr_1\\
\ni CO_r &&\labels{CO2xs}\\
\nr 0&=&2 \pi  x e^{2 \bba_0} \left(\bbm_0-\xi  x\bbm_0'\right)( \bbp_1+\bbr_1)-(\xi -1) x
   e^{ \bba_0}\bbm_0' \dot\bbu\\
\nr &-&(\xi -1)
   (x-\bbm_0) \bbm_0'
   (\dot\bbe+e^{2\bba_0}\bba_1')+4
   \pi  x^2 (x-\bbm_0) e^{2
   \bba_0}
   \bbp_1'\\
\ni EQ_{tt} &&\labels{EQ11xs}\\
\nr 0&=&\bbm_1'-4 \pi  x^2\bbr_1\\
\ni EQ_{tr} &&\labels{EQ12xs}\\
\nr 0&=&\dot\bbm_1-(\xi -1)e^{\bba_0}\bbm_0' \bbu\\
\ni EQ_{rr} &&\labels{EQ22xs}\\
\nr 0&=& \left(\xi 
  \bbm_0'-1\right)\bbm_1-4 \pi  x^2
   (x-\bbm_0) \bbp_1+2
   (x-\bbm_0)^2 (e^{-2 \bba_0} \dot\bbe+\bba_1')\\
\ni EQ_{\theta\theta} &&\labels{EQ33xs}\\
\nr 0&=&  (x-\bbm_0)  \Big(x \left(\xi \bbm_0'-2\right)+\bbm_0\Big)\bbm_1'
-2(x-\bbm_0)^2 e^{-2 \bba_0} \Big(x\left( \bbm_0'-2\right)+\bbm_0\Big) \dot\bbe\\
\nr &+&
   \bigg(\bbm_0 \Big((1-4 \xi )
  \bbm_0'-1\Big)+x \Big(\xi 
   \bbm_0'^2+(3 \xi -2)
   \bbm_0'+2\Big)\bigg)\bbm_1\\
\nr &-&2
   (x-\bbm_0)^2  \bigg(x \Big((2\xi +1)
  \bbm_0'-2\Big)-\bbm_0\bigg)\bba_1'+2 x^2 (\bbm_0-x) e^{-2 \bba_0}
   \ddot\bbm_1\\
\nr &-&16 \pi  x^2
   (x-\bbm_0)^2 \bbp_1+4 x (x-\bbm_0)^3 (e^{-2
  \bba_0} \dot\bbe'+\bba_1'')\ee 
Variables with a bar on top and a subscript 0 are functions of $x$, and  variables with a bar on top and a subscript 1, as well as $\bbu$ and $\bbe$, are functions of $x$ and $\tau$.  
A prime on a function of $(x,\tau)$ now represents $\p/\p x$, and a dot now represents $\p/\p \tau$. 

\section{Time-independent perturbation and solutions}
The first-order quantities $\bbp_1, \bbr_1, \bbm_1, \bbu, \bbe, \bba_1$ are to be solved from
 \eq{CO1xs} to \eq{EQ33xs}, but these equations still look too complicated
to be tackled directly. Fortunately, the time-independent version of these equations can be solved, and they in turn provide a stepping stone for the solution of the general time-dependent equations.

In the time-independent scenario, a small amount of matter is added to a black hole of mass $M_0$, radius $R_0$, density $\r_0(r)$, and internal mass $m_0(r)$, to obtain a black hole of mass $M_{01}=M_0+M_1:=(1+\s)M_0$, radius $R_{01}=R_0+R_1=(1+\s)R_0$, density $\r_{01}(r)=\r_0(r)+\r_1(r)$, and internal mass $m_{01}(r)=m_0(r)+m_1(r)$. We will assume $0<\s\ll 1$  to enable perturbation theory to be applied.

Since the perturbed and the unperturbed solutions are now both static, we can drop all 
the time dependences and all the time-derivative terms in \eq{CO1xs} to \eq{EQ33xs}, the $\tau$ dependence in all  functions, as well as the non-diagonal quantities $\bbu$ and $\bbe$.
As a result, \eq{CO1xs} and \eq{EQ12xs} are identically zero, and the remaining
four equations become
\be    
\ni CO_r &&\labels{CO2xs0}\\
\nr 0&=&2 \pi  x  \left(\bbm_0-\xi  x\bbm_0'\right)( \bbp_1+\bbr_1)-(\xi -1)
   (x-\bbm_0) \bbm_0'
  \bba_1'+4
   \pi  x^2 (x-\bbm_0) 
   \bbp_1'\\
\ni EQ_{tt} &&\labels{EQ11xs0}\\
\nr 0&=&\bbm_1'-4 \pi  x^2\bbr_1\\
\ni EQ_{rr} &&\labels{EQ22xs0}\\
\nr 0&=& \left(\xi 
  \bbm_0'-1\right)\bbm_1-4 \pi  x^2
   (x-\bbm_0) \bbp_1+2
   (x-\bbm_0)^2 \bba_1'\\
\ni EQ_{\theta\theta} &&\labels{EQ33xs0}\\
\nr 0&=&  (x-\bbm_0)  \Big(x \left(\xi \bbm_0'-2\right)+\bbm_0\Big)\bbm_1'\\
\nr &+&
   \bigg(\bbm_0 \Big((1-4 \xi )
  \bbm_0'-1\Big)+x \Big(\xi 
   \bbm_0'^2+(3 \xi -2)
   \bbm_0'+2\Big)\bigg)\bbm_1\\
\nr &-&2
   (x-\bbm_0)^2  \bigg(x \Big((2\xi +1)
  \bbm_0'-2\Big)-\bbm_0\bigg)\bba_1'\\
\nr &-&16 \pi  x^2
   (x-\bbm_0)^2 \bbp_1+4 x (x-\bbm_0)^3 \bba_1''\ee 

These equations still look very formidable, but fortunately their solutions can be deduced 
from the scaled solutions of Sec II.
Remember in this regard that a subscript 0 has now been added to the quantities in that section.

According to Sec.~II, the scaled density $\ol\r_0(x)$ and the scaled internal mass $\ol m_0(x)$ obey the differential equations
\eq{dless1} and \eq{dless2}, whose numerical solution for $\xi=0.16$ is given in Fig.~1 as an illustration. These scaled variables are related
to the actual variables by  $r=xR$, $\r(r)=(M/R^3)\ol\r(x)=\ol\r(x)/(2GR^2)$, and $m(r)=\ol m(x)M=\ol m(x) (R/2G)$. Thus
\be
\nr \r_0(r)&=&\ol\r_0(x_0)/2GR_0^2,\qquad m_0(r)=\ol m_0(x_0)R_0/2G,\qquad x_0=r/R_0\\
\nr \r_{01}(r)&=&\ol\r_0(x_{01})/2GR_{01}^2,\qquad m_{01}(r)=\ol m_0(x_{01})R_{01}/2G,\qquad x_{01}=r/R_{01}.\labels{rsp}\ee
Note that $\bar\r_{01}$ has been replaced by $\ol\r_0$ and $\ol m_{01}$ has been replaced by $\ol m_0$. This is allowed because the zeroth order scaled functions are universal, independent of $R$ and $M$ except through their arguments $x$.

In order to have a uniform notation later, we shall replace all $x_0$ by $x$ from now on.
With $R_{01}=(1+\s)R_0$, and $\s\ll 1$, one gets $x_{01}\simeq x(1-\s)$. Hence
\be
\ol\r  (x_{01})&\simeq& \ol\r_0  (x  )-\s x  {\ol\r}_0 '(x  ),\\
\ol m  (x_{01})&\simeq&\ol m_0  (x  )-\s x  \ol m_0'(x  ),\\
\r_{01}(r)&\simeq& {\ol\r_0  (x  )-\s x  \ol\r_0'(x  )\over 2GR_0  ^2(1+\s)^2}
\simeq\r_0  (r)-{\s\over 2GR_0  ^2}\(2\ol\r_0  (x  )+x  \ol\r_0'(x  )\),\\
m_{01}(r)&\simeq&\(\ol m_0  (x  )-2\s x  \ol m_0'(x  )\)(1+\s){R_0  \over 2G}=m_0  (r)+{\s R_0  \over 2G}\(\ol m_0  (x  )-x  \ol m_0'(x  )\).\ee

  The scaled perturbed density and internal mass are then 
\be
\ol\r_1(x)&\equiv&\(\r_{01}(r)-\r  (r)\) 2GR_0  ^2=-\s\(2\ol\r_0  (x)+x\ol\r_0'(x)\)=\-\s\bbm''_0(x)/4\pi x,\labels{s01}\\
\ol m_1(x)&\equiv&\(m_{01}(r)-m  (r)\){2G\over R_0  }=\s\(\ol m_0  (x)-x\ol m_0'(x)\),\labels{s02}\ee
Since $\bbm_0(1)=M_0$ and $\bbm_0'(1)=0$, the perturbed mass added to the original
black hole of mass $M_0$ is $\bbm_1(1)=\s M_0=M_1$, which is the expected value.

The equation of state $p(x)=-\xi\r(x)$ valid for both the unperturbed black hole and the perturbed black hole implies 
\be \bbp_1(x)=-\xi\bbr_1(x).\labels{s03}\ee
 Also, from \eq{dless3},  
\be \nr\bba_1'(x)&=&{\xi\over 1-\xi}\({1\over\bbr_0}{d\bbr_1(x)\over dx}-{\bbr_1(x)\over\bbr_0(x)^2}{d\bbr_0(x)\over dx}\)\\
&=&-{\xi\s\bigg(\-x\bbm_0''(x)^2+\bbm_0'(x)\Big(\bbm_0''(x)+x\bbm_0'''(x)\Big)\bigg)\over (1-\xi) \bbm_0'(x)^2}.
\labels{s04}\ee

 Another way to deduce \eq{s04} is to start from the scaled version of \eq{e2a}, which reads $\bba(x)=\ln\bbr(x)/2\g$. Hence 
 \be \bba_1(x)={1\over 2\g}\ln\bigg({\bbr_0(x)+\bbr_1(x)\over\bbr_0(x)}\bigg)\simeq {\bbr_1(x)\over2\g\bbr_0(x)}=-{\s\xi\over 1-\xi}{x\bbm_0''(x)\over \bbm_0'(x)},\labels{s04a}\ee
whose $x$-derivative then gives \eq{s04}.

The higher derivatives of $\bbm_0(x)$ appearing in \eq{s01} to \eq{s04a} can be converted into functions of $\bbm_0'(x)$
and $\bbm_0(x)$ by using \eq{dless1} and \eq{dless2}. For example,
\be
\nr\bbm_0''(x)&=&\frac{\bbm_0'(x) \Big(\xi  x \Big((\xi -1)\bbm_0'(x)+4\Big)+(1-5 \xi )\bbm_0(x)\Big)}{2 \xi  x (x-\bbm_0)},\\
\nr\bbm_0'''(x)&=&\frac{\bbm'_0(x)}{4 \xi ^2 x^2 (x-\bbm_0(x))^2}  \bigg( 2 \xi  x^2 \Big((\xi -1) \xi ^2 \bbm_0'(x)^2+\left(5
   \xi ^2-6 \xi +1\right) \bbm_0'(x)+4 \xi \Big)\\
   &-&\xi 
   (5 \xi -1) x \bbm_0(x) \Big(3 (\xi -1)
   \bbm_0'(x)+4\Big)+\left(15 \xi ^2-8 \xi +1\right)
   \bbm_0(x)^2\bigg).
\labels{m0hd}\ee
Substituting \eq{m0hd} into \eq{s04}, one gets 
\be \bba_1'(x)&=&\frac{\sigma  \bigg(\bbm_0(x) \big((1-7 \xi ) \bbm_0'(x)+2\big)+x \bbm_0'(x)
   \big((\xi +1) \xi  \bbm_0'(x)+4 \xi -2\big)\bigg)}{4 (x-\bbm_0(x))^2}.\labels{s05}\ee

It can be checked, with the help of \eq{m0hd},  that \eq{s01} to \eq{s04} are indeed the solutions of   \eq{CO2xs0} to \eq{EQ33xs0}.  

Before leaving this section, let us look at the signs of various quantities
and their physical implications. To start with,
the injected mass is expected to increase both $\bbr(x)$ and $\bbm(x)$, causing
 $\bbr_1(x)$ and $\bbm_1(x)$ to be positive. This can be seen analytically
 to be the case
from \eq{s01} and \eq{s02},  taking into account the behavior of $\bbr_0(x)$
and $\bbm_0(x)$ discussed in Sec.~II.  The property $\bbm_0''(x)<0\ (\Rw\bbr_1(x)>0)$ is a consequence of the decrease
of $\bbm_0'(x)$ from $\sim \b x^{\b-1}$ at small $x$, towards 0 at $x=1$. It can be seen from the convexity of $\bbm_0(x)$ in Fig.~1, as well as the exact expression of $\bbm_0''(x)$ in \eq{m0hd}. The property
 $\bbm_0(x)-x\bbm_0'(x)>0\ (\Rw\bbm_1(x)>0)$ results from the increase of 
 $\bbm_0(x)-x\bbm_0'(x)\sim x^\b(1-\b)>0$ at small $x$ towards 1 at $x=1$,
 staying positive all the way through. 
 
 Since $\xi<1$ and $\bbm_0'(x)>0$, it also follows that 
 $\bbm_0(x)-x\xi\bbm_0'(x)>0$ and $\bbm_0(x)-x\xi^2\bbm_0'(x)>0$.
 The former is needed in \eq{dless1} to keep $d\bbr_0(x)/dx$ negative and
 to ensure that $\bbr_0(x)$ decreases steadily from infinity to 0.
 Both of these two inequalities will be used later to study the behavior
 of time-dependent solutions.

\section{Solution of the first-order time-dependent equations}
Returning now to the general case, where  matter is injected into a black hole
of mass $M_0$ and radius $R_0=2GM_0$ over a period of time. In that case
$M(t):=M_0(1+\s(\tau))$, with $\s(\t)\ll 1$ at all time
so that perturbation theory can be used. The task now is to find the solution
of  the six time-dependent first-order equations \eq{CO1xs} to \eq{EQ33xs}.
Fortunately, with the help of the solution for the time-independent solutions
obtained in the last section, 
solutions of the time-dependent equations can also be found.

Since \eq{s01}, \eq{s02}, and \eq{s04a}  are true for any $\s$, we might expect their time-dependent solutions to be simply
\be
\bbr_1(x,\t)&=&-\s(\t)\(2\ol\r_0  (x)+x\ol\r_0'(x)\)=-\s(\t)\bbm_0''(x)/4\pi x
:=\hat\r_1(x)\s(\t),\labels{s01t}\\
\bbm_1(x,\t)&=&\s(\t)\(\ol m_0  (x)-x\ol m_0'(x)\):=\hat m_1(x)\s(\t),\labels{s02t}\\
 \bba_1(x,\t)&=& {\bbr_1(x,\t)\over2\g\bbr_0(x)}=-{\s(\t)\xi\over 1-\xi}{x\bbm_0''(x)\over \bbm_0'(x)}:=\hat\a_1(x)\s(\t).\labels{s04t}\ee
A priori it is possible that they may also depend on $\dot\s(\t), \ddot\s(\t)$, etc., which vanish when $\s$ is a constant.
However, the following argument shows that this does not happen, so the 
response of internal density adjustment to an external mass addition is instantaneous.

Suppose 
$ \bbm_1(x,\t)=\s(\t)\m_1(x)+\dot\s(\t)\m_2(x)$, where $ \bbm_1(x,\t)$ is the mass of a spherical ball with a scaled radius $x\le 1$ at scaled time $\t$. As such, $\bbm_1(x,\t)$ must be positive, and be a monotonically increasing function of $x$ and of $\t$ with $ \bbm_1(1,\t)=M_0\s(\t)$. Since $\s(\t)$ is arbitrary, 
at any time one can choose $\s(\t)\gg\dot\s(\t)$, or $\s(\t)\ll\dot\s(\t)$,
so both $\m_1(x)$ and $\m_2(x)$ must be  non-negative monotonically increasing functions of $x$. The condition  $ \bbm_1(1,\tau)=M_0\s(\t)$ shows that $
\m_1(1)=M_0$ and $\m_2(1)=0$. The requirement that $\m_2(x)$ be  monotonically increasing  then
shows that $\m_2(x)=0$ for all $x$. Thus $\bbm(x,\t)$ cannot depend on $\dot\s(\t)$, and similarly it cannot depend on the higher derivatives of $\s(\t)$.
Consequently \eq{s02t} should be the time-dependent solution of $\bbm_1(x,\tau)$. 

By \eq{EQ11xs}, \eq{s01t} should also be the time-dependent solution of $\bbr_1(x,\t)$, which shows that \eq{s04t} must also be true.

The rest of the variables can be obtained from \eq{CO1xs} to \eq{EQ33xs}, with the help of \eq{m0hd}, \eq{s01t}, \eq{s02t}, and \eq{s04t}.

From \eq{EQ12xs}, one gets 
\be\bbu(x,\t)=-\frac{e^{-\bba_0(x)} \big(\bbm_0(x)-x\bbm_0'(x)\big) \dot\sigma (\tau)}{(1-\xi) \bbm_0'(x)}:=\hat u(x)\dot\s(\t).\labels{ru}\ee

With \eq{ru}, and \eq{s01t},  \eq{s02t}, the right-hand side of \eq{CO1xs}
can be computed. The result is zero, so \eq{CO1xs} is satisfied,
further confirming the validity of \eq{s01t} and  \eq{s02t}.

Next, from \eq{EQ22xs} one gets 
\be
\dot\bbe &=&\frac{e^{2 \bba_0}}{4
   (x-\bbm_0)^2} \bigg(8 \pi 
   x^2 (x-\bbm_0) \bbp_1-\sigma (\tau )
   \Big(\xi  x \big((\xi -1)
  \bbm_0'+4\big)+(1-5 \xi )
  \bbm_0\Big) \bbm_0'\bigg).\labels{ret0}
   \ee
Substituting this into \eq{CO2xs}, an ordinary differential equation (in $x$) 
for $p_1(x,\t)$ emerges, 
\be \nr 0&=&16
   \pi  \xi  x^3 (x-\bbm_0)^2 e^{2
   \bba_0}
   \bbp_1'-8 \pi  \xi  x^2 (x-\bbm_0) e^{2
   \bba_0} \Big((2 \xi -1) x
  \bbm_0'-\bbm_0\Big) \bbp_1\\
\nr&&-e^{2 \bba_0}
  \bbm_0' \bigg(\xi  x^2
   \bbm_0' \Big((\xi \-1) \xi 
   \bbm_0'\-6 \xi +2\Big)\+2 \xi  x
  \bbm_0 \Big(\-\left(\xi ^2\-4 \xi
   \+1\right)  \bbm_0'\+\xi
   \+1\Big)+(1\-5 \xi )
   \bbm_0^2\bigg)\sigma (\tau
   ) \\
&&+4 \xi  x^2
   (x-\bbm_0) \left(x
    \bbm_0'-\bbm_0\right)
   \sigma ''(\tau ).
\labels{p1diff}\ee
This equation is considerably simplified if we make use  of the equation of state. In the static
case, the equation of state is $\bbp(x)=-\xi\bbr(x)$, with $\xi$ being a constant.
In the time dependent case, $\xi(x,\t):=-\bbp(x,\t)/\bbr(x,\t)$ may no longer be a constant.
If we expand $\xi(x,\t)=\xi+\xi_1(x,\t)$ perturbatively, then the first-order 
equation of state becomes 
\be \bbp_1(x,\t)=-\xi\bbr_1(x,\t)-\xi_1(x,\t)\bbr_0(x).\labels{p1xi}\ee
The differential equation obtained from \eq{p1diff} and \eq{p1xi} for $\xi_1(x,\t)$
is considerably simpler, 
\be
\nr 0&=&\xi_1(x,\t)'-f(x)\xi_1(x,\t)-g(x)\ddot\s(\t),\\
\nr f(x)&=&\frac{\bbm_0(x)-\xi ^2 x
  \bbm_0'(x)}{2 \xi  x
   (\bbm_0(x)-x)},\\
g(x)&=&\frac{x e^{-2 \bba_0(x)}
   \left(\bbm_0(x)-x
   \bbm_0'(x)\right)}{(\bbm_0(x)-x)
   \bbm_0'(x)}.\labels{xidiff}\ee
   
The use of $\xi_1$ not only simplifies \eq{p1diff}, it also reduces \eq{ret0}
to the simpler form 
\be \dot\bbe(x,\t)=\frac{e^{2 \bba_0(x)}
  \bbm_0'(x) \xi_1(x,\tau
   )}{2 (\bbm_0(x)-x)}.\labels{ret}\ee

Substituting \eq{ret0} into \eq{EQ33xs} and using \eq{p1xi}, one also gets
a differential equation for $\xi_1(x,\t)$, which turns out to be identical to the one
shown in \eq{xidiff}. This shows that \eq{EQ33xs} is not independent of the other
five first-order equations. 

To summarize, the solution of the six equations \eq{CO1xs} to \eq{EQ33xs}
is given by \eq{s01}, \eq{s02}, \eq{s04a}, \eq{p1xi}, and \eq{ret}, with $\xi_1(x,\t)$
satisfying the differential equation \eq{xidiff}.

To solve \eq{xidiff}, we need to specify boundary conditions. First of all, 
it follows from \eq{ru} that $\bbu(0,\t)=0$ (see also the discussion in the next section).
With no flow at the origin, the equation of state there must remain static,
hence $\xi_1(0,\t)=0$. Secondly, for static perturbation when the driving term
$g(x)\ddot\s(\t)=0$ in \eq{xidiff}, we must also have $\xi_1(x,\t)=0$. With these
conditions, the solution of \eq{xidiff} must be of the form $\xi_1(x,\t)=\hat\xi_1(x)\ddot\s(\t)$,
with
\be
\hat\xi_1(x)=h(x)\int_0^x{g(y)\over h(y)}dy,\labels{xis}\ee
where $h(x)$ is the solution of the homogeneous differential equation
$h'(x)-f(x)h(x)=0$, whose solution is 
\be h(x)=h_0\exp\(\int_a^x f(z)dz\).\labels{hs}\ee
Different lower limit $a$ corresponds to a different amplitude $h_0$, but it can be
seen from \eq{xis} that the solution of $\hat\xi_1(x)$ is completely independent
of the amplitude, so the choice of $h_0$ and $a$ is immaterial.

\section{Summary, discussions, and numerical illustrations}
\subsection{Summary}
When matter enters a black hole, its character is altered.  Most of its global quantum numbers such as
isospin and baryonic number are lost. What then is the property of this new kind of `black hole matter'?  Being inside a black hole, that cannot be   directly measured,  
but it must still satisfy the
Einstein equations and the covariant conservation laws, from which
 some  property can be deduced theoretically. 

In the previous paper \cite{Lam}, we found that  matter inside a neutral,
static, spherical, and non-rotational black hole  can be modelled
by an ideal fluid, whose pressure $p_0(r)$ and density $\r_0(r)$ 
obey an equation of state $p_0(r)=-\xi\r_0(r)$ for some constant $\xi>0$. 
The negative pressure present is needed to counter gravitational attraction to
allow the black hole matter to fill the interior  of the black hole.
From the
Einstein equations and the covariant conservation laws, we found that a solution
exists when the constant $\xi$ is confined to a narrow range between $1/7=0.1429$
and $3-2\sqrt{2}=0.1716$. Within this range, the internal density
$\r_0(r)$ which vanishes at the event horizon to satisfy  continuity, rises steadily and becomes divergent at the origin as demanded by the singularity theorem.

In the present article,  the dynamical response of this matter is probed  by injecting
additional matter into the black hole over a period of time. The time-dependent metric and the energy-momentum tensor for this situation are given in \eq{ds2t} and \eq{SEMT}. The resulting Einstein equations and  covariant
conservation laws are listed in Apendix A. These equations are so complicated and so non-linear
that there is no hope to solve them exactly, so we  resort to perturbation theory.  The unperturbed quantities come from the initial
static black hole with mass $M_0$, and the perturbed quantities arise from the injected mass $M_0\s(\t)$. 
The total black hole mass grows from $M_0$
to  $M(t)=M_0(1+\s(\t))$, with $0\le \s(\t)\ll 1$
at all time to allow perturbation theory to be used.

With the perturbation
expansion given  in \eq{expansion}, the  resulting six first-order Einstein equations and conservation laws are  shown in \eq{CO1xs} to \eq{EQ33xs}, where
$x=r/R_0$ is the scaled radial distance and $\t=t/R_0$ is the scaled time.
A prime represents $\p/\p x$, and a dot indicates $\p/\p\t$.

The solution of these six equations can be obtained and is shown in \eq{s01t}, \eq{s02t}, \eq{s04t}, \eq{ru},
\eq{ret}, and \eq{xis}. For convenience, they are reproduced below.
\be
\bbr_1(x,\t)&=&-\s(\t)\(2\ol\r_0  (x)+x\ol\r_0'(x)\)=-\s(\t)\bbm_0''(x)/4\pi x
:=\hat\r_1(x)\s(\t),\labels{s1s}\\
\bbm_1(x,\t)&=&\s(\t)\(\ol m_0  (x)-x\ol m_0'(x)\):=\hat m_1(x)\s(\t),\labels{s2s}\\
 \bba_1(x,\t)&=& {\bbr_1(x,\t)\over2\g\bbr_0(x)}=-{\s(\t)\xi\over 1-\xi}{x\bbm_0''(x)\over \bbm_0'(x)}:=\hat\a_1(x)\s(\t),\labels{s3s}\\
 \bbu(x,\t)&=&-\frac{e^{-\bba_0(x)} \big(\bbm_0(x)-x\bbm_0'(x)\big) \dot\sigma (\tau)}{(1-\xi) \bbm_0'(x)}:=\hat u(x)\dot\s(\t),\labels{s4s}\\
\dot\bbe(x,\t)&=&\frac{e^{2 \bba_0(x)} \bbm_0'(x) \hat\xi_1(x)\s''(\tau )}{2 (\bbm_0(x)-x)}:=\hat \e(x)\ddot\s(\t) ,\labels{s5s}\\
\nr\xi_1(x,\t)&=&\hat\xi_1(x)\ddot\s(\t),\qquad \hat\xi_1(x)=h(x)\int_0^x{g(y)\over h(y)}dy,\\
\nr && h(x)=h_0\exp\(\int_a^x f(z)dz\)\\
\nr  && f(x)=\frac{\bbm_0(x)-\xi ^2 x\bbm_0'(x)}{2 \xi  x(\bbm_0(x)-x)},\\
&& g(x)=\frac{x e^{-2 \bba_0(x)}\left(\bbm_0(x)-x\bbm_0'(x)\right)}{(\bbm_0(x)-x)
   \bbm_0'(x)},\labels{s6s}\ee
where $\xi_1(x,\t)$ is related to $\bbp_1(x,\t)$ through \eq{p1xi}.

\subsection{Discussions and limitations}
The three quantities $\bbm_1(x,\t), \bba_1(x,\t)$, and $\bbe(x,\t)$ describe how
the spacetime metric is modified. The other three quantities $\bbr_1(x,\t), \bbu(x,\t)$, and $\xi_1(x,\t)$ tell us how the energy-momentum tensor is affected by
the injection of matter, namely, the dynamical response of black hole matter.
In each of the six quantities, the spatial ($x$)  and the temporal ($\t$)
dependences factorize. The temporal function of $\bbm_1(x,\t), \bba_1(x,\t),
\bbr_1(x,\t)$ is $\s(\t)$, so the response to  an external injection of matter is instantaneous. Both $\bbu(x,\t)$ and $\bbe(x,\t)$ are proportional to $\dot\s(\t)$,
causing the non-diagonal elements of the metric and the momentum tensor to vanish
for time-independent perturbations when $\dot\s(\t)=0$. Modification 
of the static equation of state is encoded in $\xi_1(x,\t)$, which is proportional to $\ddot\s(\t)$,
so the static equation of state $\bbp_1(x,\t)=-\xi\bbr_1(x,\t)$ remains valid if 
external matter is fed in at a constant rate, when $\ddot\s(\t)=0$. How the static
equation of state is modified under an accelerated injection of matter will be
discussed below.

The $x$ dependences of these six quantities are given by functions with a hat 
on top, {\it e.g.}, $\bbr_1(x,\t)=\hat\r_1(x)\s(\t)$. It can be seen from \eq{s1s} to \eq{s6s}, and the inequalities discussed at the end of Sec.~IV, that
each of $\hat\r_1(x)$, $\hat m_1(x)$, $\hat\a_1(x)$, $-\hat u(x)$, $\hat\e(x)$, $f(x)$, $g(x)$, and $\hat\xi_1(x)$
is positive in the range $x\in(0,1)$.

Physically, the positivity of $\hat\r_1(x)$ and $\hat m_1(x)$ come about because
matter is added to the black hole. This injection causes matter to flow
inward towards, giving rise to $-\hat u(x)>0$.
 The positivity of $\hat\xi_1(x)$ is the most interesting consequence,
showing an increase in stiffness of black hole matter to counter an accelerated
inward flow, as we shall explain later.


To better understand  these first-order hat functions, we need to
know their behavior near the boundaries $x=0$ and $x=1$.
Using the behavior of the zeroth-order functions near the boundary  given in Sec.~II,
and summarized in the first few rows of Table 1 and Table 2, the behavior of
the first-order functions can  be deduced from \eq{s1s} to \eq{s6s} and is
displayed in the remaining rows of Table 1 and Table 2. Coefficients are not given except for
the function $f(x)$, because its coefficient affects the $x$-power of $h(x)$ via the
homogeneous differential equation $h'(x)-f(x)h(x)=0$.

$$\ba{|c||c|c|}\hline
&x\sim 0&x\sim 1\\ \hline\hline
\bbr_0(x)&x^{\b-3}&(1-x)^\g\\
\bbm_0(x)&x^\b&1\\
\bbm_0'(x)&x^{\b-1}&(1-x)^\g\\
\bbm_0''(x)&x^{\b-2}&(1-x)^{\g-1}\\
e^{2\bba_0(x)}&x^{-2\zeta}&(1-x)\\
\hat\r_1(x)&x^{\b-3}&(1-x)^{\g-1}\\
\hat m_1(x)&x^\b&1\\
\hat\a_1(x)&1&(1-x)^{-1}\\
\hat u(x)&x^{\zeta+1}&(1-x)^{-\h-\g}\\
f(x)&\simeq \w/x&\simeq 1/2\xi(1-x)\\
h(x)&x^\w&(1-x)^{1/2\xi}\\
g(x)&x^{\d-1}&(1-x)^{-2-\g}\\
\hat\xi_1(x)&x^{\d}&(1-x)^{-\g-1}\\ 
\hat\e(x)&x^{2-\b}&1\\
\hat\r_1(x)/\bbr_0(x)&1&1/(1-x)\\
\hat m_1(x)/\bbm_0(x)&1&1\\
\hat\a_1(x)/\bba_0(x)&1/|\ln(x)|&1/(1-x)\ln|(1-x)|\\
\hline\ea$$
\bc Table 1.\quad Boundary behavior of various functions of $x$. Coefficients
are not shown except for $f(x)$. To emphasize that point a $\simeq$
symbol is added to that row\ec
\vs
$$\ba{|c|c|c|c|}\hline
&{\rm definition}&{\rm as\ a\ function\ of}\ \xi&\xi=0.16\\ \hline
\g&\g&(1-\xi)/2\xi&2.625\\
\b&\b&(-1+7\xi)/\xi(1+\xi)&0.646552\\
\zeta&(3-\b)/2\g&(1-3\xi)/(1+\xi)&0.448276\\
\w&(1-\b\xi^2)/2\xi&(1+2\xi-7\xi^2)/2\xi(1+\xi)&3.07328\\
\d&(3-\b)(1+\g^{-1})&-3+1/\xi&2.94143\\
\hline\ea$$\bc Table 2.\quad Various parameters (column 1) as functions of $\xi$ (column 2)\\
and their values at $\xi=0.16$\ec
\vs

Shown in Table 1 are also the ratios 
$\hat\r_1(x)/\bbr_0(x)$, $\hat m_1(x)/\bbm_0(x)$,
$\hat\a_1(x)/\bba_0(x)$, because they affect
where perturbation
theory is valid. For perturbation theory to
make sense, the first-order quantities in \eq{expansion} must be much smaller than
the corresponding zeroth order quantities, and that we must also have $|\bbu(x,\t)|\ll1$, $|\bbe(x,\t)|\ll1$, and $|\xi_1(x,t)|\ll \xi$. If we examine Table 1, we see that perturbation
theory would be valid provided $\s(\t), \dot\s(\t)$, and $|\ddot\s(\t)|$ are small enough, and provided we stay far enough away from $x=1$ because of
the divergence of $\hat\r_1(x)/\bbr_0(x),\ \hat\a_1(x)/\bba_0(x),\ \hat u(x)$, and $\hat\xi_1(x)$ at $x=1$. How close we can get near $x=1$ depend on how small
$\s(\t), \dot\s(\t)$, and $|\ddot\s(\t)|$ are.

The physical origin of the divergence of $\hat\r_1(x)/\bbr_0(x)$ 
is not hard to understand. Boundary condition specifies that $\r_0(r)=0$
at $r=R$, but $\r(r)=\r_0(r)+\r_1(r)$ vanishes only at $r=R_0(1+\s)$.
Hence the ratio of $\r_1(r)/\r_0(r)$ is infinite at $r=R_0$, or equivalently $\hat\r_1(x)/\bbr_0(x)$ is infinite at $x=1$.

The divergence of $\hat u(x)$ comes about because the internal density adjustment to
an addition of external matter  is instantaneous, both proportional to
$\s(\t)$. By pushing a finite amount of matter across the event horizon
at a very small amount of time (actually 0), the flow rate at the event horizon has
to be infinite.

The divergence of $\hat\xi_1(x)$ and its implication will be discussed in the next subsection.

\subsection{Static and dynamic property of black hole matter}
$\xi_1(x,\t)=0$ if $\ddot\s(\t)=0$, so the static equation of state $\bbp(x,\t)=-\xi\bbr(x,\t)$ remains unchanged when  $\dot\s(\t)$ is a constant, in which case the flow velocity $\bbu(x,\t)$  is also a constant at every $x$. More generally,  $\xi_1(x,\t)$
and $\dot\bbu(x,\t)$ have the same time dependence $\ddot\s(\t)$, suggesting a relation between
the two. Their $x$ dependence
both grow steadily from 0 to infinity at the event horizon, so that
the modification $\xi_1(x,\t)$ becomes larger at those $x$ when the acceleration $\dot\bbu(x,\t)$ is bigger. One might therefore understand the change $\xi_1(x,\t)$
to be caused by the acceleration $\dot\bbu(x,\t)$,  to produce more (less)
negative pressure to counter the inward (outward) force brought on by the acceleration (deceleration) of the
injected matter.

That the equation of state can be affected by external conditions is also seen
in ordinary matter. 
 The constant
$\xi$ for non-relativistic matter is $\xi=0$, but if heat is added to cause the molecules
to move relativistically, then  $\xi=-\frac{1}{3}$.  If the temperature is low near the center
of a spherical system and is high near its boundary at $r=R_0$, then $\xi$ would increase steadily
as $r$ increases. This is analogous to the behavior of the dark hole matter, but with
temperature replacing  dark hole fluid acceleration.

\subsection{Numerical illustration}
The zeroth order quantities $\bbr_0(x)$ and $\bbm_0(x)$ for $\xi=0.16$
are displayed in Fig.~1. Various first-order quantities for the same $\xi=0.16$
are displayed in Figs.~2-6 to illustrate the discussions of the previous sections.

\bc\igwg{14}{Fig2}\\ Fig.~2.\quad  for $\xi=0.16$\ec
\bc\igwg{14}{Fig3}\\ Fig.~3.\quad  for $\xi=0.16$\ec
\bc\igwg{14}{Fig4}\\ Fig.~4.\quad  for $\xi=0.16$\ec
\bc\igwg{14}{Fig5}\\ Fig.~5.\quad  for $\xi=0.16$\ec
\bc\igwg{14}{Fig6}\\ Fig.~6.\quad  for $\xi=0.16$\ec

\newpage
\appendix
\section{Einstein equations and the covariant conservation laws}
The computed results of 
the covariant conservation law $CO_\n:=\nabla^\m T_{\m\n}=0$ 
and the Einstein equation $EQ_{\m\n}=G_{\m\n}-8\pi G T_{\m\n}=0$
obtained from \eq{ds2t} and \eq{SEMT} are displayed below. 
Arguments $(r,t)$ of functions are not shown, a dot represents the time derivative $\p/\p t$, and a prime represents the spatial derivative $\p/\p r$.
Only non-zero components are listed.  
$EQ_{\f\f}$ is also not listed because it is the same as  $EQ_{\theta\theta}$.
\be
\ni CO_t&&\\ 
\nr 0&=&r \rho   \bigg(G \Big(\dot m r^2-m  \epsilon   r+(r-2 Gm ) \epsilon   m'
r+2 G m ^2 \epsilon  \Big)-r(r-2 G m )^2 \epsilon' \bigg) u ^2\\
\nr &+&(r-2 Gm ) \Bigg(r (r-2 G m )\big((r-2 G m ) \epsilon^2+e^{2 \alpha  } r\big) \rho  
   (u^0)' \\
\nr &+&u^0  \bigg(e^{2 \alpha  } r \Big(\rho\big(G m'  r+3 (r-2 Gm ) \alpha'  r+2 r-5 Gm \big)+r (r-2 G m )\left(p' +\rho' \right)\Big)\\
\nr &+&\epsilon\Big(\epsilon   \left(2 \rho+r \left(p' +\rho'\right)\right) (r-2 Gm )^2+r \rho   \left(\epsilon'  (r-2 G m )^2+2 G r\dot m\right)\Big)\bigg)\Bigg) u\\
\nr &+&r (r-2 G m )\Bigg(\bigg(\epsilon   \Big(2 \rho\dot \epsilon  +\epsilon \left(\dot p +\dot \rho \right)\Big) (r-2 Gm )^2+e^{2 \alpha  } \Big(r(r-2 G m ) \left(\dot p +\dot \rho \right)\\
\nr &+&\rho  \big(\epsilon   \alpha' (r-2 G m )^2+2 r\dot \alpha   (r-2 G m )+G r\dot m \big)\Big)\bigg)(u^0) ^2\\
\nr &+&(r-2 G m ) \left((r-2 G m ) \epsilon^2+e^{2 \alpha  } r\right) \rho\left(2
  \dot {(u^0)} +u' \right) u^0+(r-2 G m ) \left((r-2 G m ) \epsilon  p' -r\dot p\right)\Bigg)\\
\nr  &+&p  \Bigg(r \bigg(G \Big(\dot m  r^2-m  \epsilon   r+(r-2 G m ) \epsilon   m' 
 r+2 G m ^2 \epsilon  \Big)-r(r-2 G m )^2 \epsilon' \bigg) u ^2\\
 \nr &+&(r-2 G m ) \bigg(r (r-2 G m ) \Big((r-2 G m ) \epsilon^2+e^{2 \alpha  } r\Big)(u^0)'\\
\nr &+&u^0  \Big(e^{2 \alpha  } r \big(Gm'  r+3 (r-2 G m ) \alpha' r+2 r-5 Gm \big)\\
 \nr &+&\epsilon   \big(2 G\dot m  r^2+(r-2 G m )^2\epsilon'  r+2 (r-2 G
   m )^2 \epsilon\big)\Big)\bigg) u\\
 \nr &+&r (r-2 G m ) u^0  \Big((r-2 G m ) \left((r-2 G m ) \epsilon^2+e^{2 \alpha  } r\right) \left(2\dot {(u^0)} +u' \right)\\
\nr &+&u^0  \left(2 \epsilon\dot \epsilon   (r-2 Gm )^2+e^{2 \alpha  }\left(\epsilon   \alpha'  (r-2 G m )^2+2 r\dot \alpha  (r-2 G m )+G r\dot m \right)\right)\Big)\Bigg)\ee
\be
\ni CO_r&&\labels{A2}\\ 
\nr 0&=&\Bigg(e^{2 \alpha  } r \bigg(\rho  \Big(2 \left(G m'  r+r-3 Gm \right)+r (r-2 G m ) \alpha ' \Big)+r (r-2 G m ) \left(p' +\rho' \right)\bigg)\\
\nr &+&\epsilon\left(\epsilon   \left(2 \rho+r \left(p' +\rho ' \right)\right) (r-2 Gm )^2+r \rho   \left(2 (r-2 Gm )^2 \epsilon' -G r\dot m \right)\right)\Bigg)
u ^2\\
\nr &+&r \Bigg((r-2 G m )\Big((r-2 G m ) \epsilon^2+e^{2 \alpha  } r\Big) \rho\left(\dot{(u^0)} +2u' \right)\\
\nr &+&u^0 \bigg(\epsilon   \Big(\rho  \dot \epsilon +\epsilon  \left(\dot p +\dot \rho \right)\Big) (r-2 Gm )^2+e^{2 \alpha  } \Big(r(r-2 G m ) \left(\dot p+\dot \rho
   \right)\\
\nr &+&\rho   \left(-2\epsilon   \alpha' (r-2G m )^2+r \dot \alpha   (r-2G m )+3 G r
 m^{(0,1)} \right)\Big)\bigg)\Bigg) u\\
\nr &+&r (r-2 G m ) \Bigg(e^{4\alpha  } (r-2 G m ) \rho  \alpha ' (u^0) ^2+(r-2 G m ) \epsilon \left(\dot p +u^0 \epsilon   \rho  \dot u \right)\\
\nr &+&e^{2 \alpha  }\left(u^0  \rho   \left(r\dot u -(r-2 G m )u^0  \left(\epsilon  
\dot \alpha  -\dot \epsilon\right)\right)+(r-2 G m )p' \right)\Bigg)\\
\nr &+&p \Bigg(\bigg(e^{2 \alpha  } r \Big(2\left(G m'  r+r-3 Gm \right)+r (r-2 G m ) \alpha'\Big)\\
\nr &+&\epsilon   \Big(2 \epsilon   (r-2 Gm )^2+r \big(2 (r-2 G m )^2
 \epsilon' -G r\dot m \big)\Big)\bigg)u ^2\\
\nr &+&r \Bigg((r-2 G m )\Big((r-2 G m ) \epsilon^2+e^{2 \alpha  } r\Big)
 \left(\dot {(u^0)} +2u' \right)\\
\nr &+&u^0\bigg(3 e^{2 \alpha  } G r\dot m +(r-2 G m ) \Big(e^{2\alpha  } r\dot  \alpha
 +(r-2 G m ) \epsilon\left(\dot \epsilon  -2e^{2 \alpha  } \alpha^{(1,0)} \right)\Big)\bigg)\Bigg) u \\
\nr &+&r (r-2 G m ) u^0 \left(\left((r-2 G m ) \epsilon^2+e^{2 \alpha  } r\right)
\dot u+e^{2 \alpha  } (r-2 Gm ) u^0  \left(-\epsilon\dot \alpha +\dot \epsilon
+e^{2 \alpha  } \alpha'\right)\right)\Bigg)\\
\ni EQ_{tt}&&\labels{A3}\\ 
\nr 0&=&e^{4 \alpha  } \epsilon   \bigg(\epsilon   \Big(4 G^2m ^2+2 r \alpha '  (r-2 G m )^2-r^2\Big)-2 G r^2\dot m -2 r \epsilon'  (r-2 G m )^2\bigg)\\
\nr &-&2G r^2 m'  e^{6 \alpha}-e^{2 \alpha  }\epsilon  ^2 (r-2 G m )^2
\left(2 r \dot \alpha   \epsilon-2 r \dot \epsilon-\epsilon  ^2\right)\\
\nr &-&8 \pi  G r^2 \Big(\epsilon^2 (r-2 G m )+r e^{2 \alpha}\Big)^2 \left((p +\rho) \left(e^{2 \alpha  }u^0 -u  \epsilon\right)^2-p  e^{2 \alpha}\right)\ee
\be
\ni EQ_{tr}&&\labels{A4}\\ 
\nr 0&=&2 G r^2 e^{4 \alpha  } \Big(r\dot m -m'  \epsilon(r-2 G m )\Big)-\epsilon
    ^5 (r-2 G m )^3\\
\nr &-&8 \pi  G r^2(r-2 G m ) \Big(\epsilon  ^2(r-2 G m )+r e^{2 \alpha}\Big)^2 \left((p +\rho) \left(u  \epsilon  -e^{2\alpha  } u^0 \right)
   \left(\frac{r u }{r-2 Gm }+u^0  \epsilon\right)+p  \epsilon\right)\\
\nr &+&e^{2 \alpha  } \epsilon(r-2 G m )^2 \bigg(-2 r\epsilon   \Big(\epsilon'  (r-2 G m )+r \dot \alpha \Big)+\epsilon  ^2\Big(2 r \alpha '  (r-2 Gm )-2 G m -r\Big)\\
\nr &+&2 r^2\dot \epsilon  \bigg)\\
\ni EQ_{rr}&&\labels{A5}\\  
\nr 0&=&-8 \pi  G r
    \Big(\epsilon  ^2
   (r-2 G m )+r e^{2 \alpha
    }\Big)^2 \bigg((p +\rho
    ) \Big(r u +(r-2 G m )u^0  \epsilon
    \Big)^2+r p \bigg)\\
\nr &+&e^{2 \alpha  }
   \bigg(-2 r^2 \epsilon   \Big(2 G
   \dot m +\dot \alpha   (r-2
   G m )\Big)\\
\nr &+&\epsilon  ^2 (r-2 G
   m ) \Big(2 G r m' +4 r
   \alpha '  (r-2 G m )-4 G
   m -r\Big)+2 r^2 \dot \epsilon  (r-2 G
   m )\bigg)\\
\nr &+&\epsilon  ^3 (r-2 G
   m ) \Big(-2 G r \dot m +2
   \epsilon '  (r-2 G
   m )^2-\epsilon   (r-2 G
   m )\Big)+2 r e^{4 \alpha  }
   \Big(r \alpha '  (r-2 G
   m )-G m \Big)\\
\ni EQ_{\theta\theta}&&\labels{A6}\\ 
\nr 0&=&-\epsilon   \dot \epsilon
    \epsilon ' 
   r^5+\epsilon  ^2 \dot \epsilon'
    r^5+G \epsilon  
   \dot m \dot \epsilon 
   r^4-G \epsilon  ^2 \ddot m 
   r^4+8 G m  \epsilon   \dot \epsilon
    \epsilon ' 
   r^4\\
\nr &-&8 G m  \epsilon  ^2 \dot \epsilon'
     r^4-4 G^2 \epsilon  ^2
   \dot m ^2 r^3-4 G^2 m 
   \epsilon   \dot m \dot  \epsilon
   r^3+4 G^2 m  \epsilon
    ^2 \ddot m r^3-24 G^2
   m ^2 \epsilon  \dot  \epsilon
     \epsilon' 
   r^3\\
\nr &+&24 G^2 m ^2 \epsilon  ^2
   \dot \epsilon'  r^3+8 G^3 m 
   \epsilon  ^2  \dot m^2 r^2+4
   G^3 m ^2 \epsilon  
   \dot m  \dot \epsilon 
   r^2-4 G^3 m ^2 \epsilon  ^2
   \ddot m  r^2\\
\nr &+&32 G^3 m ^3
   \epsilon   \dot \epsilon 
   \epsilon'  r^2-32 G^3
   m ^3 \epsilon  ^2 \dot \epsilon'
     r^2-16 G^4 m ^4
   \epsilon  \dot  \epsilon  
   \epsilon'   r+16 G^4 m ^4
   \epsilon  ^2 \dot \epsilon' 
   r\\
\nr &+&e^{4 \alpha  } (r-2 G m )^2
   \bigg((r-2 G m ) \alpha''
     r^2+\left(r \alpha'
    +1\right) \Big(G
   \left(m -r m' \right)+r
   (r-2 G m ) \alpha'
   \Big)\bigg)\\
\nr &+&e^{2 \alpha
    } \bigg(\epsilon  ^2 \Big(2
   \alpha' \left(r \alpha'
   +1\right)+r \alpha''
   \Big) (r-2 G
   m )^4-\epsilon  
   \Big(\left(r \alpha'
    +1\right) \epsilon'
     (r-2 G m )^2\\
\nr &+&r
   \dot \alpha  \left(G
 m' r-r+G m \right)-G r
   \dot m  \left(r \alpha'
  +1\right)\Big) (r-2 G
   m )^2\\
\nr &+&G r^3 \dot m  \dot \alpha
    (r-2 G m )+r \Big(r
   \left(-\left(\dot \alpha  
   \epsilon'  -\dot \epsilon'
  \right) (r-2 G
   m )^2-G r
   \ddot m \right)\\
\nr &-&(r-2 G m )
  \dot  \epsilon  \big(G m +r
   \left(G
   m' -1\right)\big)\Big)
   (r-2 G m )-3 G^2 r^3
   \dot  m ^2\bigg)\\
\nr &-&8 G \pi 
   r (r-2 G m )^2 p  \Big((r-2 G
   m ) \epsilon  ^2+e^{2 \alpha
    } r\Big)^2
   \ee

\end{document}